\documentclass[structabstract]{aa}  
\usepackage{graphicx,natbib}
\bibpunct{(}{)}{;}{a}{}{,}
\usepackage{txfonts}

\newcommand\msol{\ensuremath{\,\mbox{\it M}_{\odot}}}

\newcommand\lta{\mathrel{\hbox{\raise 0.6 ex \hbox{$<$}\kern
                   -1.8 ex\lower .5 ex\hbox{$\sim$}}}}
\newcommand\gta{\mathrel{\hbox{\raise 0.6 ex \hbox{$>$}\kern
                   -1.7 ex\lower .5 ex\hbox{$\sim$}}}}
 
\newcommand{\scrbox}[1]{\ensuremath{{\mbox{\scriptsize #1}}}}

\newcommand{\teff}{{\ensuremath{T_{\scrbox{eff}}}}}
\newcommand{\Msol}{\ensuremath{\,\mbox{\it M}_{\odot}}}
\newcommand{\Mstar}{\ensuremath{\it \,M_{*}}}

\newcommand{\Dturb}{\ensuremath{D_{\scrbox{T}}}}

\newcommand{\MS}{main--sequence}
\newcommand{\gr}{\ensuremath{g_{\scrbox{rad}}}}

\newcommand{\DM}{\ensuremath{ \log \Delta M/M_{*}}}

\newcommand{\He}{\mbox{He}}

\newcommand{\Fe}{\mbox{Fe}}
\newcommand{\Mn}{\mbox{Mn}}
\newcommand{\Mg}{\mbox{Mg}}

\newcommand{\Cr}{\mbox{Cr}}
\newcommand{\Ca}{\mbox{Ca}}
\newcommand{\Ni}{\mbox{Ni}}

\newcommand{\Ti}{\mbox{Ti}}

\newcommand{\Li}{\mbox{Li}}

\begin{document}
   \title{Sirius\,A: turbulence or mass loss?}

   \author{G. Michaud \inst{1,2} \and J. Richer
          \inst{2}
          \and
          M. Vick\inst{2}
          }

   \institute{LUTH, Observatoire de Paris, CNRS, Universit\'e Paris Diderot,
     5 Place Jules Janssen,
     92190 Meudon, FRANCE
     \and  
   D\'epartement de Physique, Universit\'e de Montr\'eal,
       Montr\'eal, PQ, H3C~3J7, CANADA\\
              \email{michaudg@astro.umontreal.ca,jacques.richer@umontreal.ca,mathieu.vick@umontreal.ca}   
                  }

   \date{\today}

 
  \abstract
   {Abundance anomalies  observed in a  fraction of A and B stars of both Pop I and II are apparently related to internal particle transport.}
   {Using available constraints from Sirius\,A, we wish to determine how well evolutionary models including atomic diffusion  can explain observed abundance anomalies when either turbulence or mass loss is used as the main competitor to atomic diffusion.  }
   {Complete stellar evolution models, including the effects of atomic diffusion and radiative accelerations, have been computed from the zero age \MS{} of 2.1\msol{} stars for metallicities of $Z_0= 0.01$\,$\pm 0.001$ and shown to agree with the observed parameters of Sirius\,A.  Surface abundances were predicted for three values of the mass loss rate and 
   for four values of the mixed surface zone.}
   {
   A mixed mass of $\sim 10^{-6} \msol$ \emph{or} a mass loss rate of 10$^{-13}$\,\msol/yr were determined through comparison with observations. Of the 17 abundances determined observationally which are included in our calculations, up to 15 can be predicted within  2 $\sigma$ and 3 of the 4 determined upper limits are compatible. }
   { While the abundance anomalies can be reproduced slightly better using turbulence as the process competing with atomic diffusion, mass loss probably ought to be preferred since the mass loss rate required to fit abundance anomalies is compatible with the observationally determined rate.  A mass loss rate within a factor of 2 of $10^{-13} \Msol/\mbox{yr}$ is preferred. This restricts the range of the directly observed mass loss rate.}


   \maketitle
%

\section{Astrophysical context}
\label{sect:context}
The brightest star in the sky, Sirius\,A, has all its main parameters such as mass, luminosity, age... relatively well determined, and its abundances have now been studied in great detail  using Space Telescope data by \citet{Landstreet2011}, where more background information may also be found.

In previous work, the then available observations of Sirius were compared with results from a grid of models calculated with atomic diffusion and turbulence as a competing process by \citet{RicherMiTu2000}.  Given the range of abundances  observed for any given species, the agreement seemed satisfactory.
In a recent paper \citep{VickMiRietal2010} similar results were obtained through a similar approach  but with mass loss as the competing process.
Because the observational error bars were too large, these authors were unable to delineate which of mass loss or turbulence was responsible for its abundance anomalies.  

It is currently  uncertain as to which process is most efficient in competing with atomic diffusion in A stars. For O  and early B stars, mass loss is most likely the dominant process.  It is clearly observed in those stars at a rate sufficient to obliterate the effects of atomic diffusion.  However in \MS{} A stars  the expected mass loss rate due to radiative accelerations is smaller than in O stars by several orders of magnitude. Its presence is likely only if it is started by another mechanism \citep{Abbott82}, and it might involve only metals \citep{Babel95}.  On the other hand, since surface convection zones are very thin, one expects little corona driven flux as observed on the Sun.  It is then \emph{a priori} quite uncertain if A stars have any mass loss and the claimed mass loss rate for Sirius is  an important observation. It is thus important to verify as precisely as possible if it is compatible with current observed abundance anomalies on the surface of Sirius.

The  measurement of the mass loss rate of Sirius is a difficult observation and awaited Hubble telescope measurements of the Mg\,II resonance lines \citep{BertinLaVietal95}. Their analysis of this  spectral feature with a wind model leads to an uncertainty of 0.5 dex on the mass loss rate if Mg is all once ionized.  However there is an additional  uncertainty related to the evaluation of the fraction of \Mg{} that is once ionized. Their more credible evaluation of Mg ionization is based on setting the ionization rate equal to the recombination rate and leads to\footnote{This is slightly different from their Eq.\,[17] for which they had neglected the error bar given with their Eq.\,[7] but included an  evaluation of the Mg II ionization based on a \emph{corrected LTE} value coming from atmosphere models \citep{SnijdersLa75} which do not seem appropriate for the wind region of interest here.}:
\begin{equation}
     6 \times 10^{-14} < - \frac{dM}{dt} < 5 \times 10^{-13} \Msol/\mbox{yr}.
\label{eq:rate}
\end{equation}

On the other hand, turbulence has often been used in stellar evolution calculations to explain observed abundance anomalies.  In F stars of clusters, turbulence could be responsible for the destruction of surface \Li{} \citep{TalonCh98,TalonCh2005} thereby leading to the so called \Li{} gap. Turbulence could also be responsible for reducing abundance anomalies caused by atomic diffusion on Am and Fm stars \citep{TalonRiMi2006}.  It can naturally explain the disappearance of abundance anomalies as rotation increases in those objects.  It could also play a role for the \Li{} abundance evolution in solar type stars \citep{PinsonneaultKaSoetal89,ProffittMi91a}.  It has however always been found necessary to use a number of adjustable parameters for its description by physical models of turbulent transport and its role is uncertain.  In this series of papers on the role of atomic diffusion in stellar evolution, turbulence was only introduced when models with atomic diffusion led to anomalies larger than observed.  Only one parameter was adjusted in order to control the influence of turbulence in limiting the size of anomalies:  the surface mixed mass \citep{RicherMiTu2000,MichaudRiRi2011}.

 It is possible to improve what we learn from the acccurate observations of Sirius by making more precise evolutionary calculations.   Instead of the grid of solar metallicity models used in \citet{RicherMiTu2000} and in \citet{VickMiRietal2010},
 this paper uses two new series of models which were respectively computed  with turbulence and with mass loss as the process competing with atomic diffusion.  These two series are precisely converged to the known properties of Sirius, $L$, \teff, $R$, $M$ and age. The original metallicity is determined and models with this metallcity are used to compare with observed abundances.  This allows for a more precise and rigorous test than the calculations using grids of models.

In this paper stellar evolution models with atomic diffusion as described in \citet{RicherMiTu2000} and in \citet{VickMiRietal2010} are used to  determine the  original metallicity of Sirius\,A using the age, radius and mass as constraints (Sect.\,\ref{sec:OriginalMetallicity}).  Using this metallicity and the determined parameters, complete evolutionary models are then calculated (Sect.\,\ref{sec:Models}), and the surface abundances are compared with Landstreet's recent observations (Sect.\,\ref{sec:SurfaceAbundances}).  The results are discussed in Sect.\,\ref{sec:Conclusion}.

\section{Original metallicity}
\label{sec:OriginalMetallicity}
   \begin{figure*}
   \centering
\includegraphics[width=18cm]{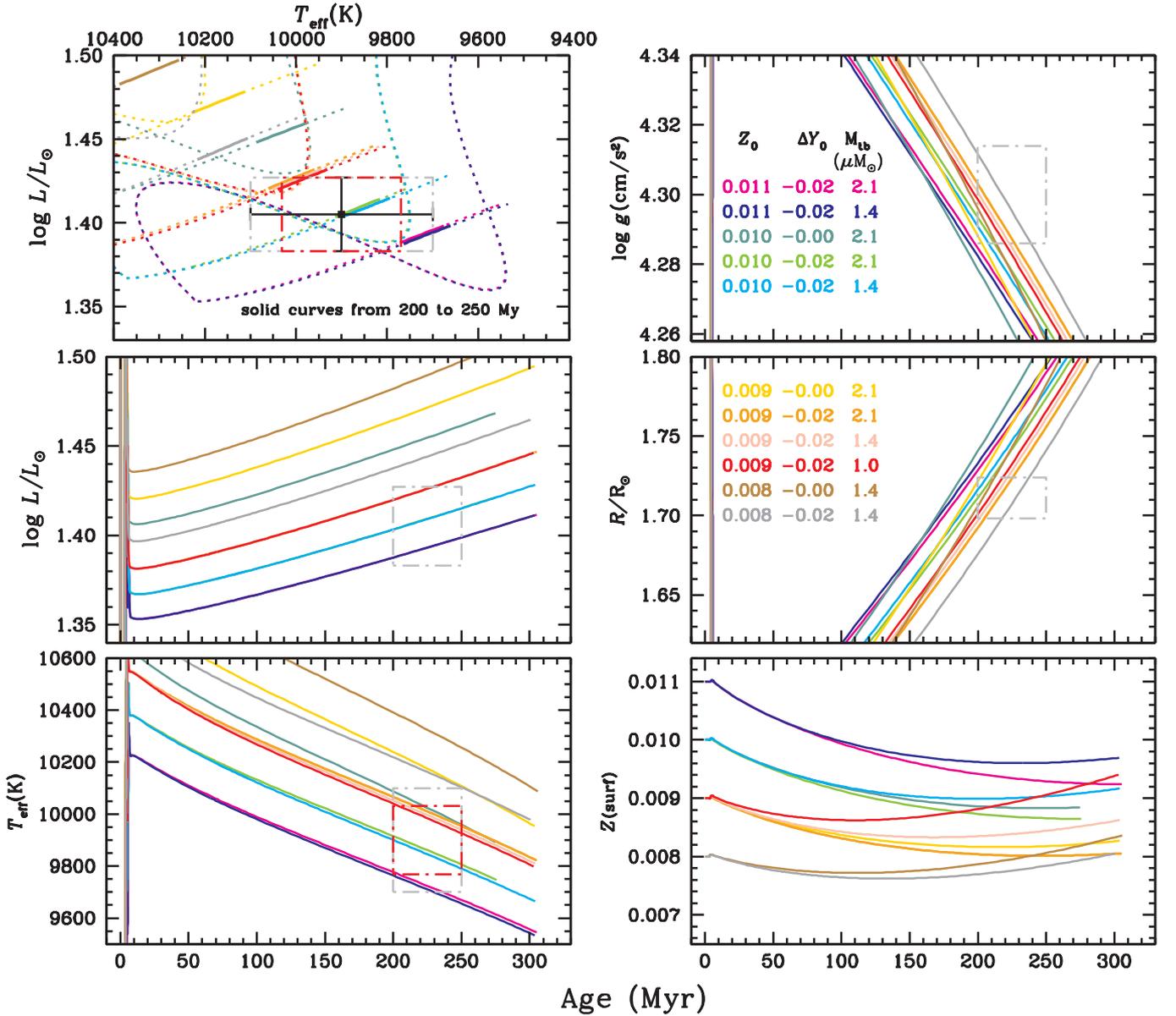}
      \caption{ An HR diagram and the time evolution of \teff{}, $L$, $\log g$, $R$ and $Z_{\rm{surf}}$ are shown. Observationally determined  $\pm 1 \sigma$ intervals are shown for  $L$, \teff{}, $g$ and  $R$.  For \teff{} the spectroscopically determined error bar is in black while that determined using luminosity and radius is in red (see the text).  Each model is color coded and identified on the figure.  The adopted acceptable age range is from 200 to 250 Myr (see the text). In the HR diagram, the part of the curves between 200 and 250 Myr is solid; it is dotted outside of that interval.  All models were calculated with turbulence.  Models with mass loss could not be distinguished, in the HR diagram, from those of the  same original mass and composition calculated with turbulence.  A subset of the curves is shown on Fig.\,\ref{fig:HR1}.}
         \label{fig:HR}
   \end{figure*}
   
The fundamental parameters required to carry out stellar evolution calculations, except for the original chemical composition, have been relatively well determined for Sirius\,A.  The Hipparcos parallax can be used to determine its distance and when coupled with interferometry \citep{KervellaThMoetal2003}, to  determine its radius, 1.711\,$\pm$0.013\,\ensuremath{\,\mbox{\it R}_{\odot}}.  These authors also use the Hipparcos parallax to determine the luminosity from the magnitude and to refine the mass determination to 2.12 $\pm$0.06 \msol{}. From the luminosity and radius one can obtain $\teff = 9900 \pm$140\,K while from spectroscopy \citet{Lemke89}  obtained $\teff = 9900 \pm$200\,K.  The age of Sirius was discussed in Sect.\,4.1 of \citet{RicherMiTu2000}; using evolutionary time scales of Sirius B, they suggested  250\,$\pm$50\,Myr.  We adopt the slightly more restrictive range 225\,$\pm$25\,Myr suggested by \citet{KervellaThMoetal2003} in their Sect.\,2.2, where they argue that Sirius B would have been more massive on the \MS{} than assumed by \citet{RicherMiTu2000}.   
Since the mass is well determined, we use a fixed mass of 2.1\msol{}, and then evolve models with a range of metallicities, using a scaled solar mix as given in Table\,1 of \citet{TurcotteRiMietal98}. Helium was adjusted  to the fitted solar value for model H of \citet{TurcotteRiMietal98} as starting homogeneous composition for some of the calculations.  For most of the calculations however, the He mass fraction was corrected by $\Delta Y =-0.02$\,dex because of the lower final metallicity\footnote{This 0.02 reduction of $Y$ for a 0.01 reduction of $Z$ is the same proportionality as used by \citealt{VandenBergSwRoetal2000} in building his Table\,2. LiBeB were taken from the meteoretic values of \citealt{GrevesseSa98}. }.   

In this paper a solar mass fraction, $\ensuremath{\,\mbox{\it X}_{\odot}}$, is often used in particular to normalize \emph{both} observed and calculated mass fractions.  They are from Table\,1 of \citet{TurcotteRiMietal98} and correspond to the solar mass fractions at the birth of the Sun, more precisely the $Z_0(= 0.01999)$ for model H in Table 6 of that paper.  	The surface solar abundances of metals today are some 10\,\% smaller.   Those normalizing factors are different from the solar abundances used by Landstreet for comparative purposes.  Since, in this paper, the same normalizing factors are used for both observed and calculated quantities, they do not influence the comparison.

Age, luminosity, \teff{} and radius are assumed well determined and are used as constraints to determine the original metallicity using models with turbulence.  In Fig.\,\ref{fig:HR},  
only models with $Z_0 = 0.009$, 0.010 and 0.011 are seen to satisfy all constraints within the predetermined error boxes.  The radius is generally satisfied only for models  younger than 250\,Myr.  Furthermore only models with the He mass fraction reduced by 0.02 satisfy the constraint on $L$.  Models with higher $Z$ and a solar He mass fraction do not satisfy the constraint from the radius as may be seen from the trend of the models with the solar He mass fraction.

Below the deepest surface
convection zone, the turbulent diffusion coefficient has been assumed
to obey a simple algebraic dependence on density given, in the calculations with turbulence 
presented in this paper, by
\begin{equation}
     \Dturb= 10^4 D(\He)_0\left(\frac{\rho_0}{\rho}\right)^4 \label{eq:DT}
\end{equation}
where  $D(\He)_0$ is the atomic diffusion 
coefficient\footnote{The values of $D(\He)_0$
actually used in this formula were always 
obtained --- for programming convenience ---
from the simple analytical approximation
\hbox{$D(\He)=3.3\times10^{-15}T^{2.5}/[4\rho\ln(1+1.125\times10^{-16}T^3/\rho)]$}
(in cgs units) for He in trace amount in an ionized hydrogen plasma. 
These can differ significantly from the more accurate values used
elsewhere in the code.} of He at some reference depth. 
Let $\Delta M \equiv \Mstar - M_r$ be the mass outside  the sphere of radius $r$.   For this paper,  
a series of models with different surface mixed masses (proportional to our  parameter $\Delta M_0$) were calculated.
More precisely calculations were carried out with 
\begin{equation}
     \rho_0=\rho(\Delta M_0).   \label{eq:Delta-M-0}
\end{equation}
where Eq.~(\ref{eq:Delta-M-0}) is given by the current stellar model.  In
words, in the calculations with turbulence reported in this paper, $\rho_0$ of Eq.~(\ref{eq:DT}) is the density found at depth
$\Delta M = \Delta M_0$ in the evolutionary model.  In practice the outer $\sim 3 \times \Delta M_0$ of the star is mixed by turbulence; for $\Delta M_0 = 10^{-6.0}$\,\Msol{} the concentration of most species is constant for $\Delta M \lta 10^{-5.5}$\,\Msol{}.    As one increases $\Delta M_0$, one  defines a one parameter family of models.

\section{The models of Sirius\,A}
\label{sec:Models}
Two series of models were evolved for Sirius\,A. One in which the process competing with diffusion is turbulence as described in \citet{RicherMiTu2000} and \citet{MichaudRiRi2011} and one in which it is mass loss as described in \citet{VickMiRietal2010}.  In both, using opacity spectra from \citet{IglesiasRo96}, all aspects of atomic diffusion transport are treated in detail from first principles.

   \begin{figure*}
   \centering
\includegraphics[width=18cm]{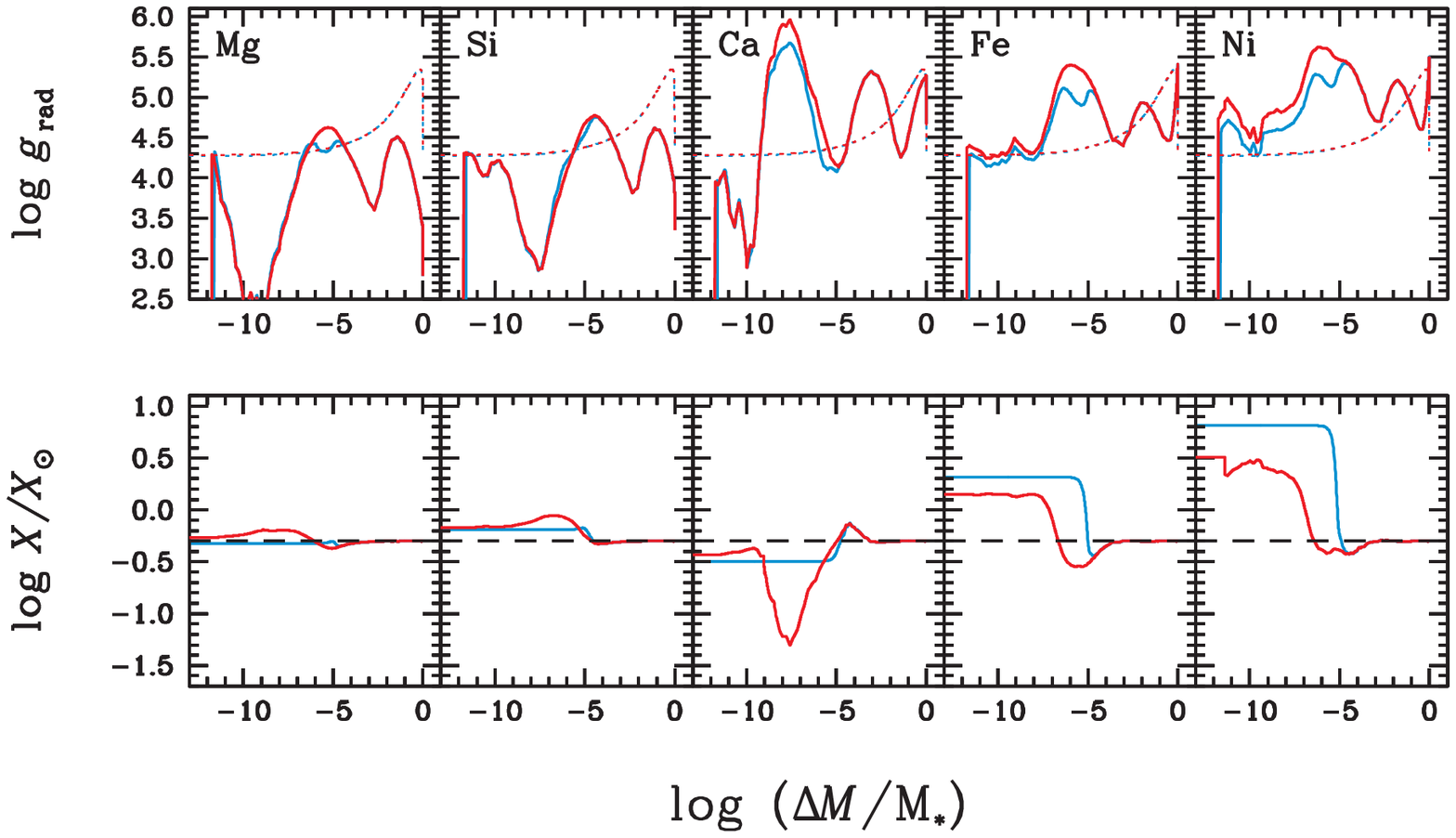}
      \caption{\emph{Upper row} Radiative accelerations for five atomic species, after 232\,Myr of evolution in a model with a mass loss rate of  10${^{-13}}$ \Msol/yr (red curves)  and in a model with turbulence (blue curves).
      The dotted lines represent gravity. \emph{Lower row} Corresponding mass fraction in the model with mass loss and in that  with turbulence. The dashed lines are the original values.  They correspond to an original metallicity of $Z_0 = 0.01$.}
         \label{fig:gR_x5}
   \end{figure*}
The radiative accelerations of Mg, Si, Ca, Fe and Ni  are shown in the upper row of Fig.\,\ref{fig:gR_x5} at $\sim$232\,Myr,  the approximate age of Sirius\,A, in both a model calculated with mass loss (red curves) and one calculated with turbulence (blues curves).  The two are practically superposed for $\DM > -5$ but are significantly different for many species closer to the surface (that is $\DM < -5$).  In the lower row the corresponding internal concentrations for the model with mass loss and for that with turbulence are shown.  When there is a difference between the \gr's for the two models they are caused by effects of saturation, as may be seen by comparing the abundances in the lower row.  The $X(\Fe)$ and $X(\Ni)$ are larger in the model with turbulence for $\DM < -5$ than in the model with mass loss; the reduction of the photon flux at the wavelengths where \Fe{} and \Ni{} absorb the most is by a larger factor when the abundance is larger   and so the \gr's are smaller in the model with turbulence.  The large underabundance of \Ca{} at $\DM \sim -7.5$ in the wind model causes the larger \gr(\Ca) there, but the large underabundance  is also caused by the maximum of \gr(\Ca) as discussed in Sec. 5.1.1 of \citet{VickMiRietal2010} to which the reader is referred for a detailed discussion of the interior wind solution.  While the surface abundances of say \Fe{} and \Ni{} in the model with mass loss are within 0.3 dex of those in the model with turbulence, their interior mass fractions differ by a factor of about 5 for $-7< \DM <-5$.

Figures \ref{fig:internal_x} and \ref{fig:gR} contain results for all calculated species; they are are shown in the online Appendix \ref{sec:Appendix}.

\section{Surface abundances }
\label{sec:SurfaceAbundances}
   \begin{figure*}
   \centering
\includegraphics[width=9cm]{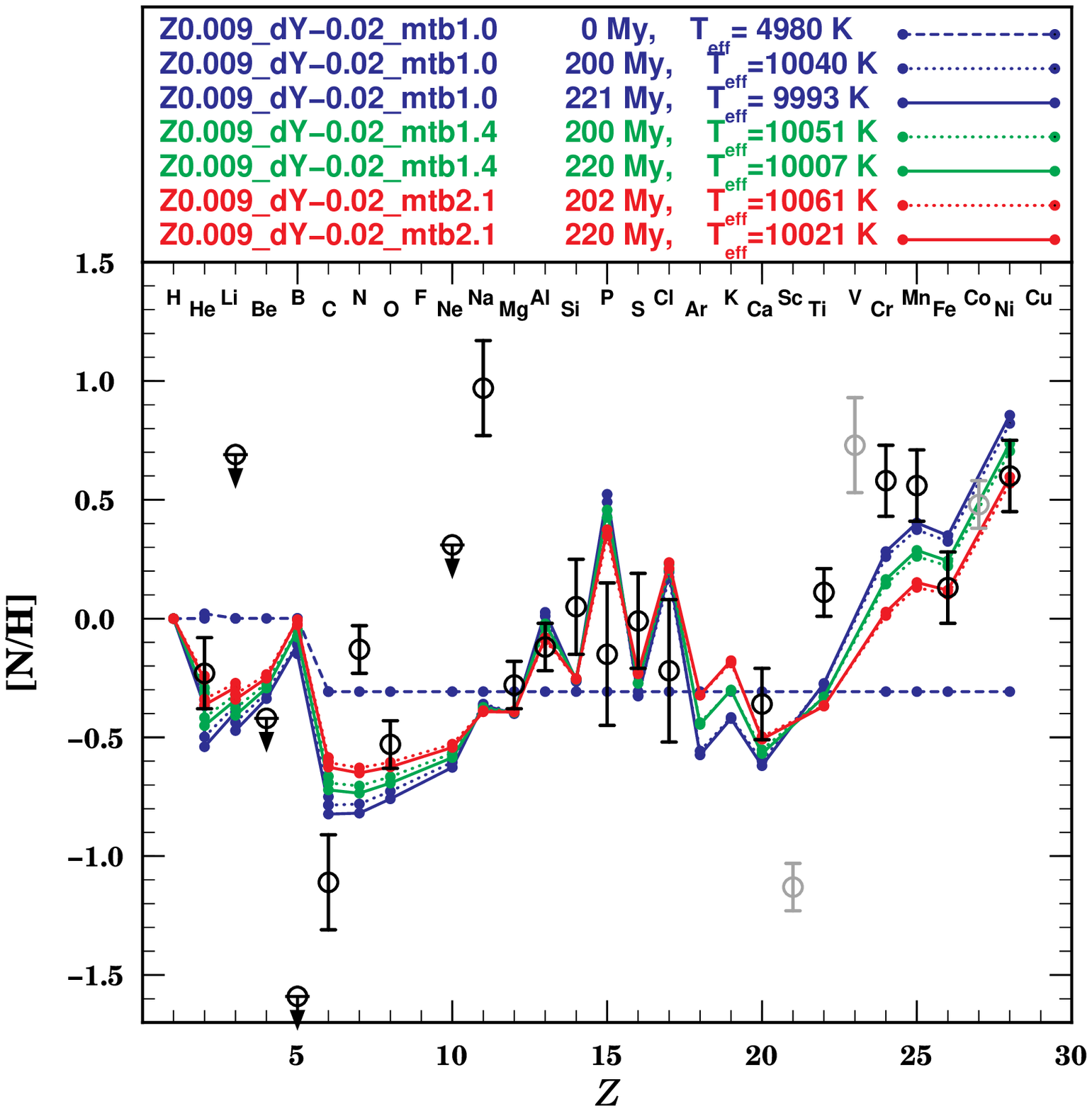}
\includegraphics[width=9cm]{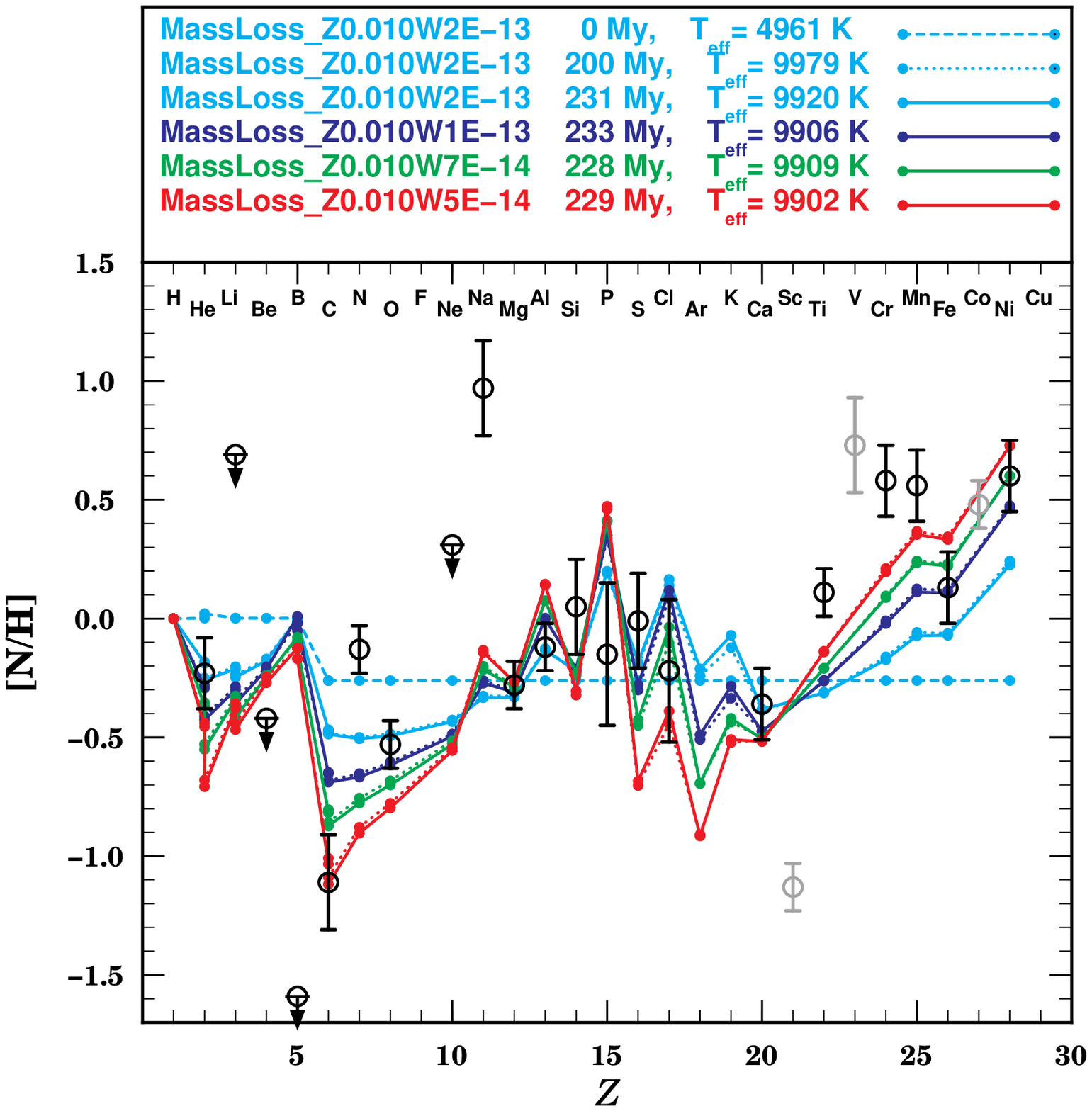}
      \caption{Observed and predicted abundances on Sirius\,A as a function of $Z$, the atomic number,   \emph{left} in the model with turbulence calculated with an original metallicity of 0.009 for three slightly different values of turbulence and \emph{right} in the model with mass loss with an original metallicity of 0.010 calculated with four different mass loss rates.  The model name Z0.009\_dY-0.02\_mtb1.0 stands for a model calculated with $Z_0= 0.009$, $\Delta Y = -0.02$, and $\Delta M_0 = 1.0 \times 10^{-6}$\,\msol.  The model name MassLoss\_Z0.010W2E-13 stands for a model calculated with  \hbox{$dM/dt = -2\times 10^{-13}$\,\msol/yr.}   All dotted lines represent models of about 200\,Myr.}
         \label{fig:surfAbun}
   \end{figure*}
   
      \begin{figure*}
   \centering
\includegraphics[width=9cm]{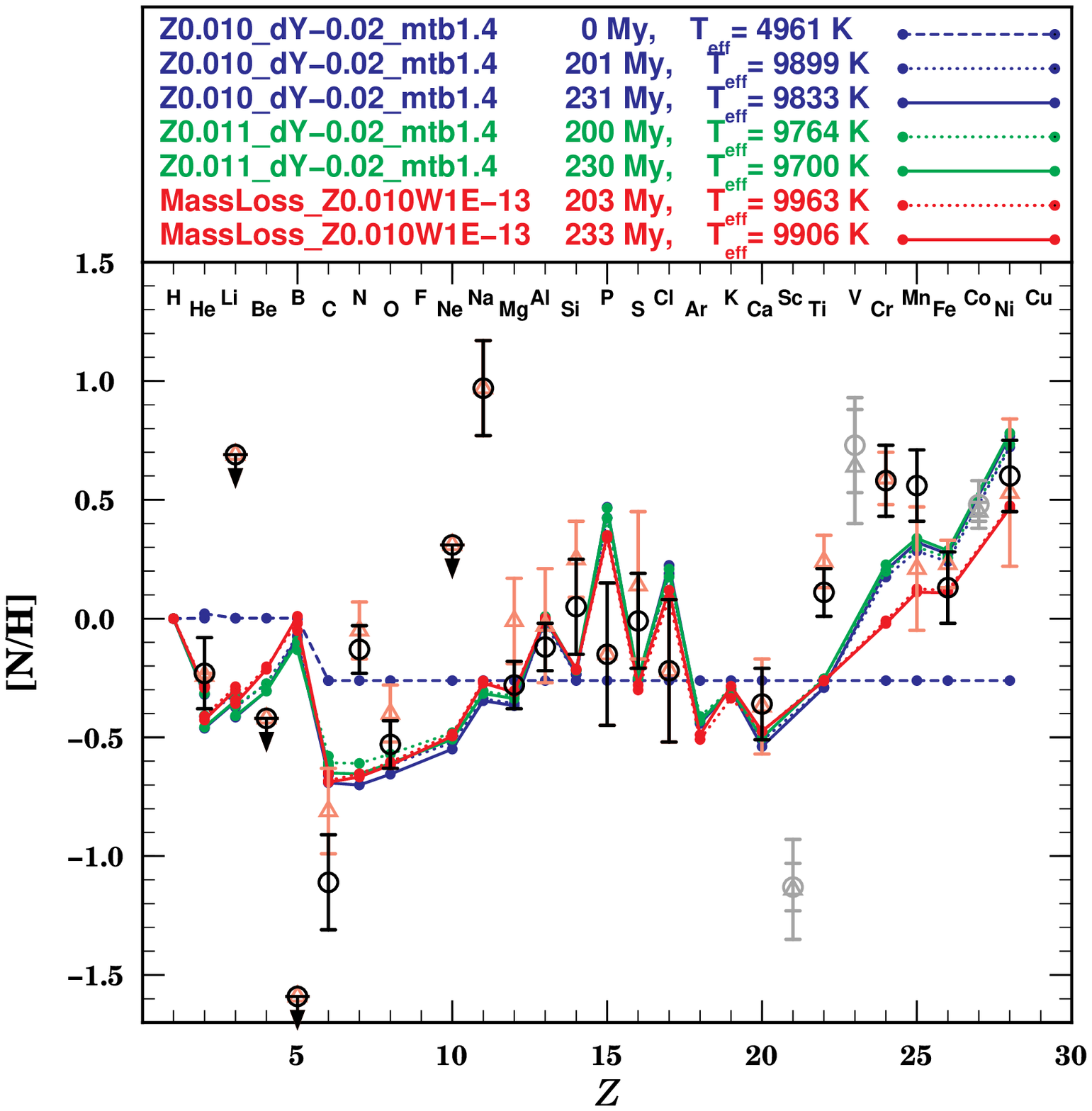}
\includegraphics[width=9cm]{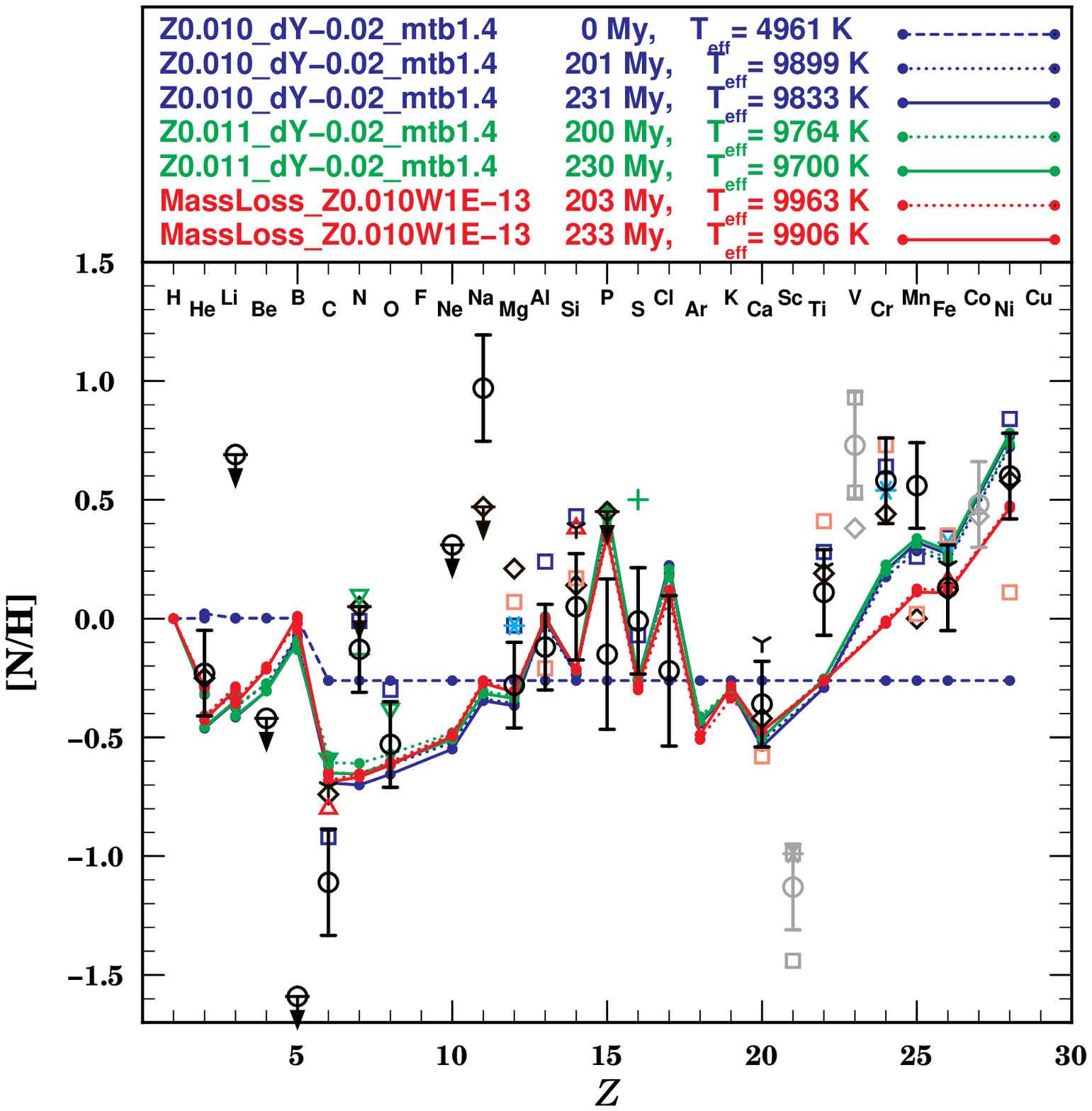}
      \caption{Observed and predicted abundances  on Sirius\,A as a function of $Z$, the atomic number, in the model with turbulence for two slightly different values of  original metallicity, 0.010 and 0.011, and in the model with a mass loss  rate of \hbox{10$^{-13}$\,\msol/yr} and a metallicity of 0.01 (see Fig.\,\ref{fig:surfAbun} for model labeling definitions).  In the \emph{left} panel,  the two sets of observations are from \citealt{Landstreet2011}: the black circles are from his observations while the pink triangles are an average over all recent observations (see the text).
      In the  \emph{right} panel, the averaged values are replaced by the actual data points of each observer as given in Table\,I of \citet{Landstreet2011} where 
{\it circles}, \citet{Landstreet2011}; 
{\it inverted open triangles}, \citet{LambertRoBe82}; 
{\it inverted three-point stars}, \citet{Lemke90};
{\it blue squares}, \citet{QiuZhChetal2001};
{\it diamonds}, \citet{HillLa93};
{\it asterisks}, \citet{HuiBonHoaBuAl97};
{\it plus}, \citet{RentzschHolm97};
{\it upright open triangles}, \citet{HolwegerSt93};
{\it pink squares}, \citet{SadakaneUe89}.
See the text for the explanation of the differences between the error bars in the right and left panels. 
 }
         \label{fig:surfAbun_turb_twin}
   \end{figure*}  
   
 Given the constraints imposed by age, \teff{}, $L$ and $R$, and that the mass is 2.1\,\msol{}, the original metallicity is fixed to \hbox{$Z_0 = 0.010 \pm 0.001$} (see Sect.\,\ref{sec:OriginalMetallicity}).  Specifically, the luminosity (middle left hand panel of Fig.\,\ref{fig:HR}) determines $Z_0 = 0.010 \pm 0.001$ and then the radius determines the acceptable age  to lie between 200 and $\lta 230$\,Myr (middle right hand panel). There only remains mass loss rates \emph{or} mass mixed by turbulence that may be varied to define a range of predicted abundances that can then be compared with observations. 
 
 Evolutionary models were calculated for $\Delta M_0= 1.0 $,  1.4 and 2.1$\times 10^{-6}\msol$ (see Eq.\,[\ref{eq:Delta-M-0}]) and for mass loss rates of 0.5, 0.7, 1.0 and 2.0 $\times 10^{-13}$\,\msol/yr.  On Fig.\,\ref{fig:surfAbun}, predictions from some of them are compared with observations  from column 2 of Table\,1 of \citet{Landstreet2011}.  On Fig.\ref{fig:surfAbun_turb_twin} data from the other columns of his Table\,1 are also used in order to present a picture of the uncertainties of the observations, as briefly discussed below.  
 
 In the left panel of Fig.\,\ref{fig:surfAbun}, results are shown for the case $Z_0= 0.009$ at 200 Myr (dotted line segments) and at 220 Myr (solid line segments) which is the  age interval over which all constraints are satisfied according to Fig.\,\ref{fig:HR}.   In practice, even though shown for all cases, the dotted and solid segments are barely distinguishable and merely widen the dots.  Note the assumed original composition at age 0.0\,Myr in light blue on each panel of Figs.\,\ref{fig:surfAbun} and \ref{fig:surfAbun_turb_twin}.
 As the mass mixed by turbulence is decreased from 2.1 to 1.0$\times 10^{-6}\msol$, the surface abundance of elements supported by \gr{} (e.g. most \Fe{} peak elements) and the underabundance factor of sinking elements both increase.   In the left panel of Fig.\,\ref{fig:surfAbun_turb_twin} similar results are shown for original metallicities of $Z_0 =0.010$ and 0.011 with  $\Delta M_0=  1.4 \times 10^{-6}\msol$.  Within the range of original metallicities acceptable according to Sect.\,\ref{sec:OriginalMetallicity} (from $Z_0 = 0.009$ to 0.011), the level of agreement between predicted surface abundances and observed ones does not change much although the $Z_0= 0.010$ case is slightly favored.  Given the number of observed species, there is in practice little room for adjustment: as one may see from  the left panel of Fig.\,\ref{fig:surfAbun}, the \Fe{} peak abundances favor the lower value of the mixed mass but the abundances of He, O, S and Ca rather favor the larger value.
 
 Predictions for four mass loss rates and a metallicity of \hbox{$Z_0= 0.010$} are shown in the right panel of Fig.\,\ref{fig:surfAbun}.  Only one metallicity is shown since the effects of changing metallicity over the acceptable range of $Z_0= 0.009$ to 0.011  are small as briefly mentioned above for the turbulence models.  The iron peak elements would probably favor a mass loss rate of 0.5 $\times 10^{-13}$\,\msol/yr, but the lower mass species then disagree more strongly, and the better compromise appears to be the 1.0 $\times 10^{-13}$\,\msol/yr case.  The iron peak elements show approximately the same sensitivity to the mass loss range as to the mixed mass range illustrated in Fig.\,\ref{fig:surfAbun} but He, O and S are  more sensitive to the mass loss rate\footnote{As the mass loss rate is reduced, the settling velocity becomes closer to the wind velocity in magnitude.  This tends to amplify differences in settling velocities  caused, among other factors, by small mass differences.  For instance  the abundance variation of  $^4\He$ is a factor of 1.8 larger than that of  $^3\He$ for the 0.5 $\times 10^{-13}$\,\msol/yr case but only a factor of 1.3 larger in the $ 10^{-13}$\,\msol/yr case (see the right panel of Fig.\,\ref{fig:surfAbun}).}.  
 
    \begin{figure*}
   \centering
\includegraphics[width=9cm]{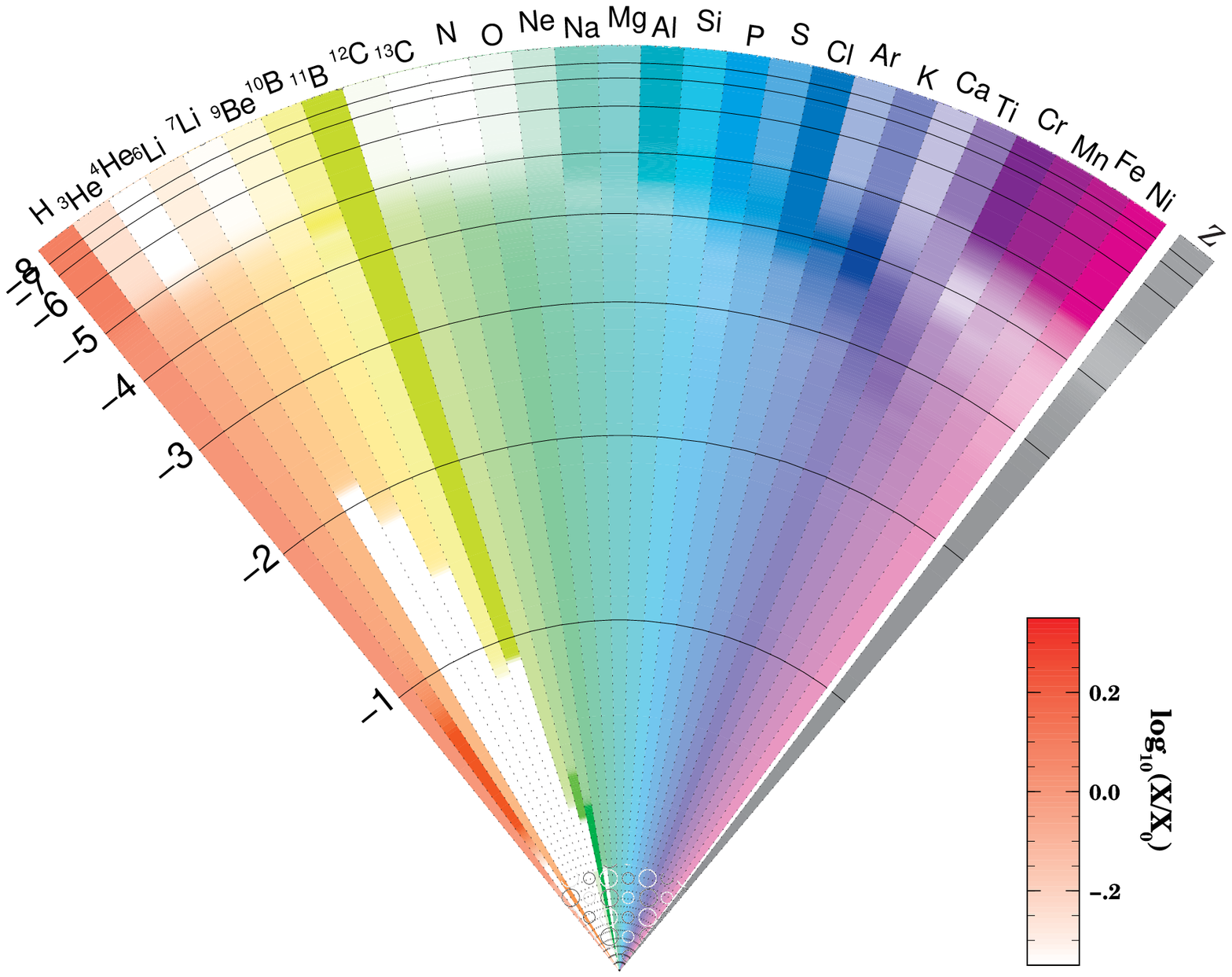}
\includegraphics[width=9cm]{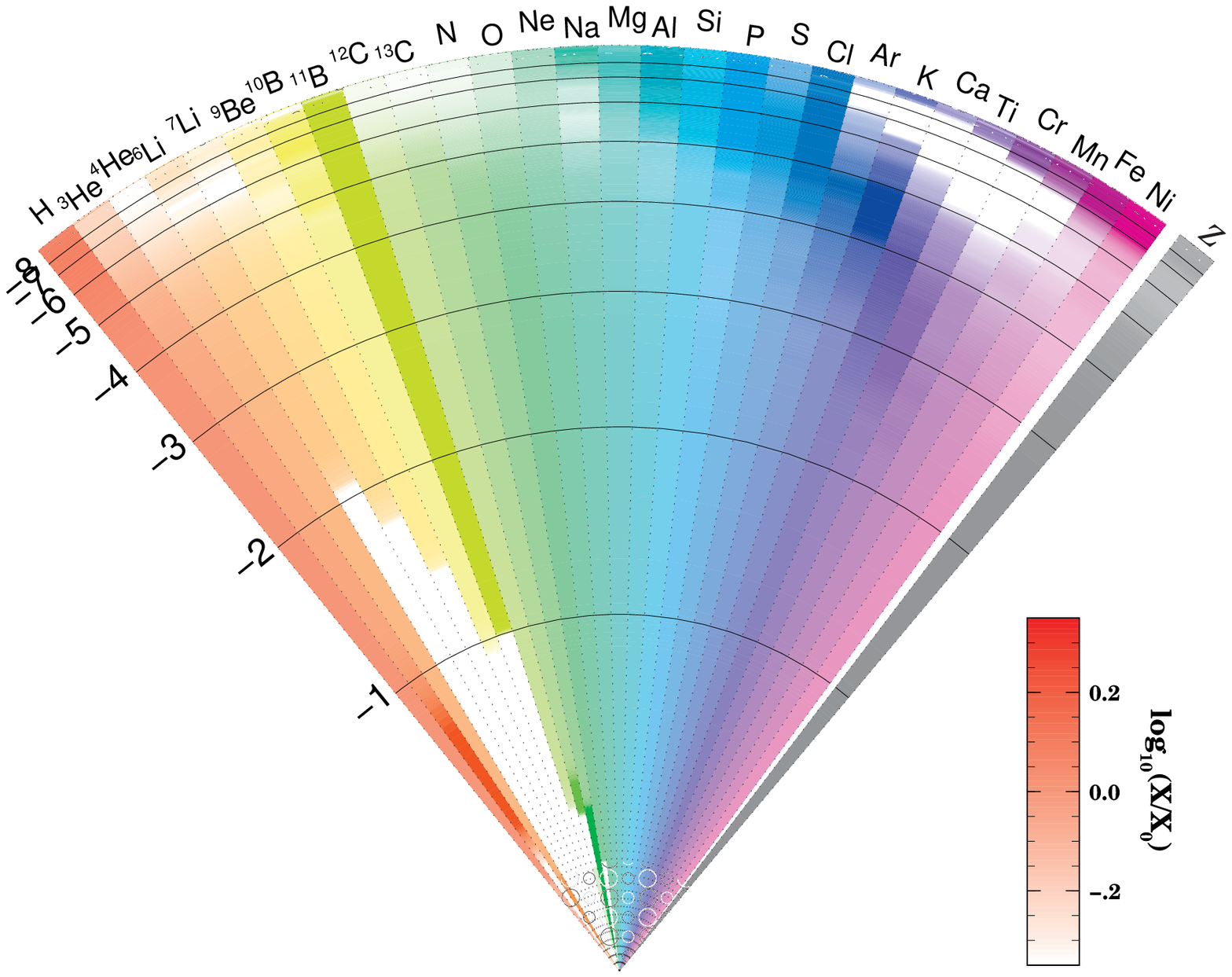}
      \caption{Color--coded interior concentrations for the same $Z_0= 0.010$ models as in Fig.\,\ref{fig:surfAbun_turb_twin} at $\sim 233$\,Myr  \emph{left} in the model with turbulence  and \emph{right} in the model with mass loss.  The radial coordinate is the radius and its scale  is linear, but the logarithmic value of the mass coordinate above a number of points, \DM, is shown on the left of the 
	horizontal black line.  The concentration scale is given in the right insert.  
	  Small  circles near the center of both models mark the central convection zone. 
      While the surface abundances are very similar, as seen in Fig.\,\ref{fig:surfAbun_turb_twin}, the interior concentrations are quite different between $\DM = -3$ and the surface. }
         \label{fig:eventails}
   \end{figure*}

   In the left and right panels of Fig.\,\ref{fig:surfAbun_turb_twin} are shown the same three theoretical models (two calculated with turbulence and one with mass loss) compared to two different presentations of the results from Table 1 of \citet{Landstreet2011}.  On the left panel are shown both his determinations (his column labeled \emph{L11}) and,  as separate data points, his determinations averaged with those of the other observers he lists in his table, except for a few values he argues are erroneous or too uncertain to be worth including (his column labeled \emph{mean}).  The error bars are standard deviations of the observations listed in his table.  These do not include contributions  from the error bars of the various authors.  The actual values of the different observers listed in his Table 1 are shown in the right panel of Fig.\,\ref{fig:surfAbun_turb_twin}.  The values which Landstreet excluded from his averages are not shown but all others are shown.  One can argue that the difference in abundance values between the various observers is a better estimate of the uncertainty of the abundance determinations
   than the mean value with associated standard deviation shown in the left panel.
   Following a discussion with Landstreet (private communication) the error bars for his points (from his column \emph{L11}) were slightly increased for the right panel of Fig.\,\ref{fig:surfAbun_turb_twin} only. So all cases where, in the "L11" column of Table 1, he had 0.1, we 
increased to 0.15 because this is closer to the actual dispersion found for dominant ions with many lines.  Noting that there is some additional uncertainty (about 0.1\,dex) due to imprecise 
fundamental parameters and microturbulence,   
 another 0.1\,dex was added in quadrature to all sigmas, giving a minimum sigma 
of 0.18.  

The right panel of Fig.\,\ref{fig:surfAbun_turb_twin} gives our  comparison between theory and observations.  The agreement is not perfect for any value of turbulence or mass loss.  In fact the difference between the mass loss and turbulent models is very small for atomic species lighter than  Cr and one may question if abundances alone can really distinguish between the two given the error bars.  Of the species included in our calculations, abundances were determined observationally for 17 atomic species and upper limits for 4.
 For the model $Z_0 =0.010$ with  $\Delta M_0=  1.4 \times 10^{-6}\msol$ one counts 8 species within 1 $\sigma$ and an additional 6 within 2 $\sigma$. These numbers are respectively 8 and 7 for the $Z_0=0.011$ model with the same turbulence.  One counts respectively 8 and 5 for the $Z_0 =0.010$ mass loss model with a mass loss rate of 1.0\,$\times 10^{-13}$\,\msol/yr.  In addition, of the 4 upper limits, 3 are compatible with predictions.  While the trend is right, the agreement is not so good for  Ti, Cr and Mn.
 
 The interior concentrations  in the two $Z_0 = 0.01$ models of Fig.\,\ref{fig:surfAbun_turb_twin} are shown in Fig.\,\ref{fig:eventails}.  While the surface abundances are quite similar in the two, the interior concentrations are quite different for $\DM < -3$ (see Sect.\,\ref{sec:Models}).  
   
   If one compares these results with  Fig.\,20 of \citet{VickMiRietal2010},  one notes that the same mass loss rate of 1.0\,$\times 10^{-13}$\,\msol/yr had been found to lead to the prediction closest to observed abundances.    While the age assumed for Sirius is about the same in the two papers, the larger mass of the models used in \citet{VickMiRietal2010} causes them to be more evolved and, so, have a smaller gravity at a similar age.  This is an important difference, when one compares two models with approximately the same \teff{} and thus the same \gr's.  Another difference comes from the original $Z_0$, which is 0.010 for the mass loss models of Fig.\,\ref{fig:surfAbun_turb_twin} of this paper, but 0.02 for those of Fig.\,20 of \citet{VickMiRietal2010}.

\section{Conclusion}
\label{sec:Conclusion}
Using observationally determined  stellar parameters for Sirius\,A one first fixed a metallicity and He mass fraction (Sect.\ref{sec:OriginalMetallicity}). Then, expected surface abundances were predicted as a function of either a surface mass mixed by turbulence or of a mass loss rate.  Of the 17 abundances determined observationally, up to 15 can be predicted within  2 $\sigma$, and 3 of the 4 determined upper limits are satisfied.  The three atomic species, B, N, and Na  show the strongest disagreement.

While the origin of the assumed turbulence is not determined, it could be either shear induced by differential rotation \citep{TalonRiMi2006} or it could be gravity waves \citep{TalonCh98}.  If the main competing process is mass loss however, it has the great advantage of being already observed \citep{BertinLaVietal95}.  The  mass loss rate of 1.0 $\times 10^{-13}$\,\msol/yr that best reproduces abundance observations is slightly larger than the lower limit of   6 $\times 10^{-14}$\,\msol/yr  determined from asymetries of Mg II lines using ST observations by these authors {(see Eq.\,[\ref{eq:rate}])}.
Their estimate based on \emph{corrected LTE}, their Eq.\,[12], however gives a mass loss rate  between 5.0 $\times 10^{-12}$ and    5.0 $\times 10^{-11}$\,\msol/yr which would lead to practically no surface abundance variations during evolution in contradiction with the observed Sirius abundances (see Figs.\,11 and 20 of \citealt{VickMiRietal2010} for a calculation with 1.0 $\times 10^{-12}$\,\msol/yr).
Our results then support the arguments presented above in Sect.\,\ref{sect:context}  in favor of their estimate based on \emph{radiative ionization fraction.} While by themselves, abundances do not favor mass loss or turbulence as the competing process, the agreement with the observationally determined mass loss rate favors mass loss.

While there probably still remains some uncertainties in the observationally determined abundances, the disagreement between observations and calculations  points to some weaknesses of the models.  The first one may be related to the presence of Sirius B.  Even though it is quite a wide pair, it had been suggested by \citet{RicherMiTu2000} that the disagreement with the C and N observations could be caused by the transfer of material from the former primary.  This was further discussed  by \citet{Landstreet2011} who also suggested that it could simultaneously explain the difficulty with the B upper limit. Sodium could also be affected, just as it is affected in the globular cluster M4 \citep{MarinoViMietal2011}.  However the extent of the mass transfer in Sirius remains uncertain.

Even if it is tempting to accept mass loss as the most important mechanism competing with diffusion in slowly rotating stars, it is also observed that abundance anomalies are much less important in rapidly rotating stars.  An other mechanism linked to rotation must then be involved.    Either rotation driven turbulence \citep{TalonRiMi2006} or meridional circulation \citep{CharbonneauMi91} could progressively reduce abundance anomalies as rotation increases.  The weak dependence of abundance anomalies on the rotation rate could be due to its effect becoming larger than those of mass loss only as one approaches the 100 km/s limit of $v \sin i$ \citep{Abt2000} for the Am star phenomenom.  

In relation to Fig.\,\ref{fig:eventails}, it was suggested in Sect.\,\ref{sec:SurfaceAbundances} that asterosismology tests could distinguish between mass loss and turbulence as the competing mechanism for slowly rotating stars.  While detailed evolutionary calculations with meridional circulation have not yet been carried out, one expects that laminar meridional circulation would lead to internal metal distributions, in 3 dimensions, similar to those that mass loss leads to, in 1 dimension, since both are advective and not diffusive processes.  This opens the possibility of distinguishing between meridional circulation and rotation induced turbulence using asterosismology.

 For simplicity, these calculations were carried out with undifferentiated mass loss throughout Sirius\,A's evolution
as seems appropriate for Am stars.  However a 2.1\,\msol{} star starts its \MS{} at $\teff \sim 10500$\,K (see Fig.\,\ref{fig:HR}) where no H convection zone is present and which is probably within the HgMn domain.  What would have been the mass loss rate then? The observation of Hg isotope anomalies on HgMn stars suggests that the mass loss would be differentiated (see \citealt{MichaudReCh74} and Sect.\,4 of \citealt{MichaudRi2008}).  How would this affect surface abundances during later evolutionary stages such as reached by Sirius\,A?

\begin{acknowledgements}
 We thank Dr John Landstreet  for very kindly communicating to us his results ahead of publication and for useful discussions. We thank Dr David Leckrone for a constructive criticism of the manuscript and useful suggestions that led to significant improvements.   This research was partially supported at  the Universit\'e de Montr\'eal 
by NSERC. We thank the R\'eseau qu\'eb\'ecois de calcul de haute
performance (RQCHP)
for providing us with the computational resources required for this
work.

\end{acknowledgements}


\begin{thebibliography}{32}
\expandafter\ifx\csname natexlab\endcsname\relax\def\natexlab#1{#1}\fi

\bibitem[{Abbott(1982)}]{Abbott82}
Abbott, D.~C. 1982, ApJ, 259, 282

\bibitem[{Abt(2000)}]{Abt2000}
Abt, H.~A. 2000, ApJ, 544, 933

\bibitem[{Babel(1995)}]{Babel95}
Babel, J. 1995, A\&A, 301, 823

\bibitem[{Bertin {et~al.}(1995)Bertin, Lamers, Vidal-Madjar, Ferlet, \&
  Lallement}]{BertinLaVietal95}
Bertin, P., Lamers, H.~J.~G.~L.~M., Vidal-Madjar, A., Ferlet, R., \& Lallement,
  R. 1995, A\&A, 302, 899

\bibitem[{Charbonneau \& Michaud(1991)}]{CharbonneauMi91}
Charbonneau, P. \& Michaud, G. 1991, ApJ, 370, 693

\bibitem[{Grevesse \& Sauval(1998)}]{GrevesseSa98}
Grevesse, N. \& Sauval, A.~J. 1998, Space~Sci.~Rev., 85, 161

\bibitem[{Hill \& Landstreet(1993)}]{HillLa93}
Hill, G.~M. \& Landstreet, J.~D. 1993, A\&A, 276, 142

\bibitem[{{Holweger} \& {Sturenburg}(1993)}]{HolwegerSt93}
{Holweger}, H. \& {Sturenburg}, S. 1993, in Astronomical Society of the Pacific
  Conference Series, Vol.~44, IAU Colloq. 138: Peculiar versus Normal Phenomena
  in A-type and Related Stars, ed. {M.~M.~Dworetsky, F.~Castelli, \&
  R.~Faraggiana}, 356--+

\bibitem[{Hui~Bon~Hoa {et~al.}(1997)Hui~Bon~Hoa, Burkhart, \&
  Alecian}]{HuiBonHoaBuAl97}
Hui~Bon~Hoa, A., Burkhart, C., \& Alecian, G. 1997, A\&A, 323, 901

\bibitem[{Iglesias \& Rogers(1996)}]{IglesiasRo96}
Iglesias, C.~A. \& Rogers, F.~J. 1996, ApJ, 464, 943

\bibitem[{{Kervella} {et~al.}(2003){Kervella}, {Th{\'e}venin}, {Morel},
  {Bord{\'e}}, \& {Di Folco}}]{KervellaThMoetal2003}
{Kervella}, P., {Th{\'e}venin}, F., {Morel}, P., {Bord{\'e}}, P., \& {Di
  Folco}, E. 2003, A\&A, 408, 681

\bibitem[{Lambert {et~al.}(1982)Lambert, Roby, \& Bell}]{LambertRoBe82}
Lambert, D.~L., Roby, S.~W., \& Bell, R.~A. 1982, ApJ, 254, 663

\bibitem[{Landstreet(2011)}]{Landstreet2011}
Landstreet, J.~D. 2011, A\&A, in press, 00

\bibitem[{{Lemke}(1989)}]{Lemke89}
{Lemke}, M. 1989, A\&A, 225, 125

\bibitem[{{Lemke}(1990)}]{Lemke90}
{Lemke}, M. 1990, A\&A, 240, 331

\bibitem[{{Marino} {et~al.}(2011){Marino}, {Villanova}, {Milone}, {Piotto},
  {Lind}, {Geisler}, \& {Stetson}}]{MarinoViMietal2011}
{Marino}, A.~F., {Villanova}, S., {Milone}, A.~P., {et~al.} 2011, ApJ, 730, L16

\bibitem[{{Michaud} {et~al.}(1974){Michaud}, {Reeves}, \&
  {Charland}}]{MichaudReCh74}
{Michaud}, G., {Reeves}, H., \& {Charland}, Y. 1974, A\&A, 37, 313

\bibitem[{{Michaud} \& {Richer}(2008)}]{MichaudRi2008}
{Michaud}, G. \& {Richer}, J. 2008, Contributions of the Astronomical
  Observatory Skalnate Pleso, 38, 103

\bibitem[{{Michaud} {et~al.}(2011){Michaud}, {Richer}, \&
  {Richard}}]{MichaudRiRi2011}
{Michaud}, G., {Richer}, J., \& {Richard}, O. 2011, A\&A, 529, A60

\bibitem[{Pinsonneault {et~al.}(1989)Pinsonneault, Kawaler, Sofia, \&
  Demarque}]{PinsonneaultKaSoetal89}
Pinsonneault, M.~H., Kawaler, S.~D., Sofia, S., \& Demarque, P. 1989, ApJ, 338,
  424

\bibitem[{Proffitt \& Michaud(1991)}]{ProffittMi91a}
Proffitt, C.~R. \& Michaud, G. 1991, ApJ, 380, 238

\bibitem[{{Qiu} {et~al.}(2001){Qiu}, {Zhao}, {Chen}, \& {Li}}]{QiuZhChetal2001}
{Qiu}, H.~M., {Zhao}, G., {Chen}, Y.~Q., \& {Li}, Z.~W. 2001, ApJ, 548, 953

\bibitem[{{Rentzsch-Holm}(1997)}]{RentzschHolm97}
{Rentzsch-Holm}, I. 1997, A\&A, 317, 178

\bibitem[{Richer {et~al.}(2000)Richer, Michaud, \& Turcotte}]{RicherMiTu2000}
Richer, J., Michaud, G., \& Turcotte, S. 2000, ApJ, 529, 338

\bibitem[{{Sadakane} \& {Ueta}(1989)}]{SadakaneUe89}
{Sadakane}, K. \& {Ueta}, M. 1989, PASJ, 41, 279

\bibitem[{{Snijders} \& {Lamers}(1975)}]{SnijdersLa75}
{Snijders}, M.~A.~J. \& {Lamers}, H.~J.~G.~L.~M. 1975, A\&A, 41, 245

\bibitem[{{Talon} \& {Charbonnel}(1998)}]{TalonCh98}
{Talon}, S. \& {Charbonnel}, C. 1998, A\&A, 335, 959

\bibitem[{{Talon} \& {Charbonnel}(2005)}]{TalonCh2005}
{Talon}, S. \& {Charbonnel}, C. 2005, A\&A, 440, 981

\bibitem[{Talon {et~al.}(2006)Talon, Richard, \& Michaud}]{TalonRiMi2006}
Talon, S., Richard, O., \& Michaud, G. 2006, ApJ, 645, 634

\bibitem[{Turcotte {et~al.}(1998)Turcotte, Richer, Michaud, Iglesias, \&
  Rogers}]{TurcotteRiMietal98}
Turcotte, S., Richer, J., Michaud, G., Iglesias, C., \& Rogers, F. 1998, ApJ,
  504, 539

\bibitem[{{VandenBerg} {et~al.}(2000){VandenBerg}, {Swenson}, {Rogers},
  {Iglesias}, \& {Alexander}}]{VandenBergSwRoetal2000}
{VandenBerg}, D.~A., {Swenson}, F.~J., {Rogers}, F.~J., {Iglesias}, C.~A., \&
  {Alexander}, D.~R. 2000, ApJ, 532, 430

\bibitem[{{Vick} {et~al.}(2010){Vick}, {Michaud}, {Richer}, \&
  {Richard}}]{VickMiRietal2010}
{Vick}, M., {Michaud}, G., {Richer}, J., \& {Richard}, O. 2010, A\&A, 521, A62

\end{thebibliography}
%
\Online

\begin{appendix} 
\section{Properties of models}
\label{sec:PropertiesOfModels}
   \begin{figure*}[h]
   \centering
\includegraphics[width=18cm]{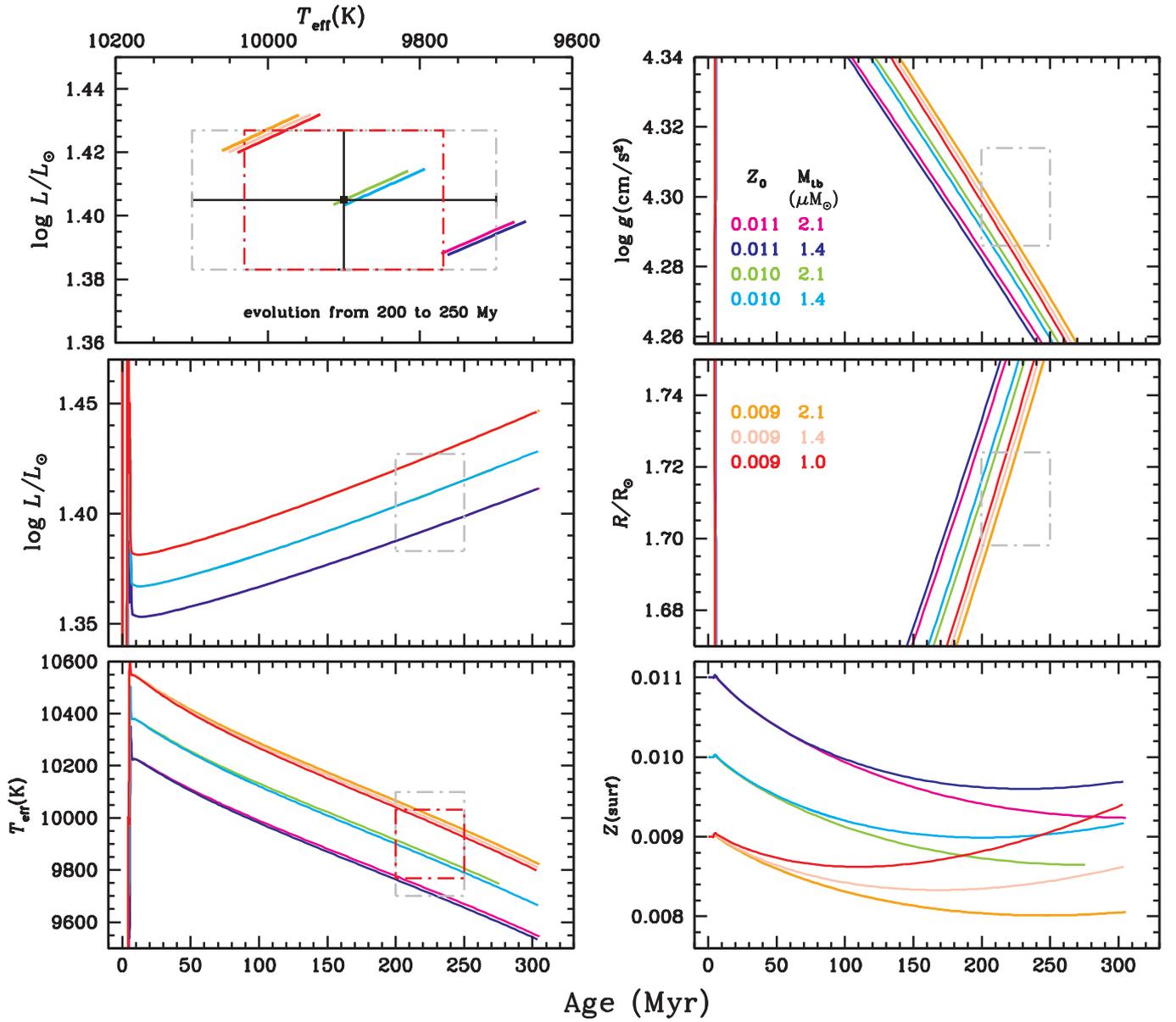}
      \caption{An HR diagram and the time evolution of \teff{}, $L$, $\log g$, $R$ and $Z_{\rm{surf}}$ are shown. Observationally determined  $\pm 1 \sigma$ intervals are shown for  $L$, \teff{}, $g$ and  $R$.  For \teff{} the spectroscopically determined error bar is in black while that determined using luminosity and radius is in red (see the text).  Each model is color coded and identified on the figure.  The adopted acceptable age range is from 200 to 250 Myr (see the text). In the HR diagram, only the part of the curves between 200 and 250 Myr is shown; see Fig.\,\ref{fig:HR} for a more complete figure.  All models were calculated with turbulence.   }
         \label{fig:HR1}
   \end{figure*}

\section{Radiative accelerations and interior concentrations for all  species }
\label{sec:Appendix}
 
   \begin{figure*}
   \centering
\includegraphics[width=18cm]{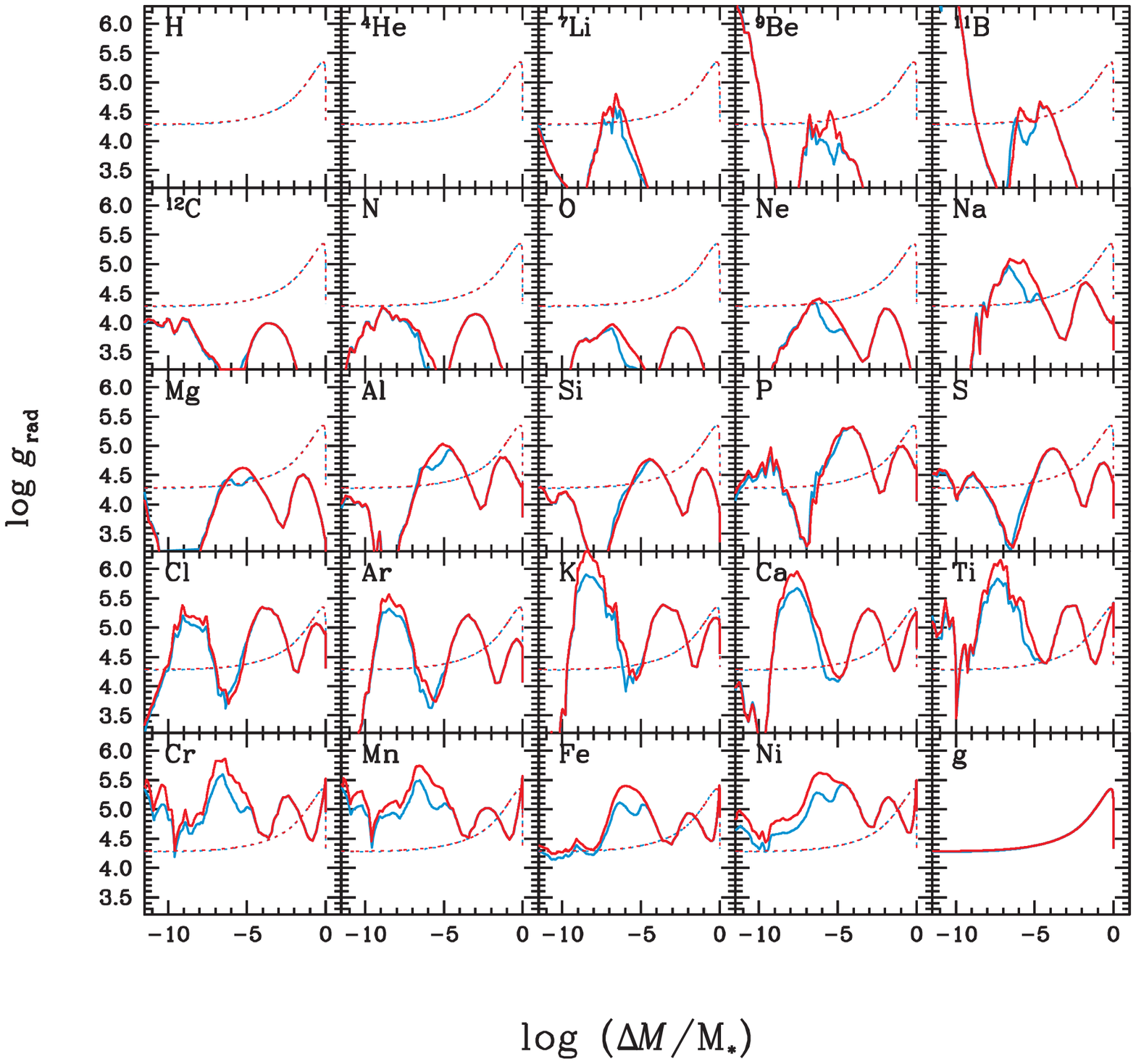}
      \caption{ Radiative accelerations after 232\,Myr evolution in a model with mass loss of  10$^{-13}$ \Msol/yr (red curves) and in a model with turbulence with $\Delta M_0=  1.4 \times 10^{-6}\msol$ (blue curves) for all calculated atomic species.  The sharp mimima in the \gr(\Ti), \gr(\Cr), and \gr(\Mn) curves at $\DM \sim -10$ appear surprising but have been verified to be a systematic property of the OPAL atomic data.}
         \label{fig:gR}
   \end{figure*}
   
 \begin{figure*}
   \centering
\includegraphics[width=16cm]{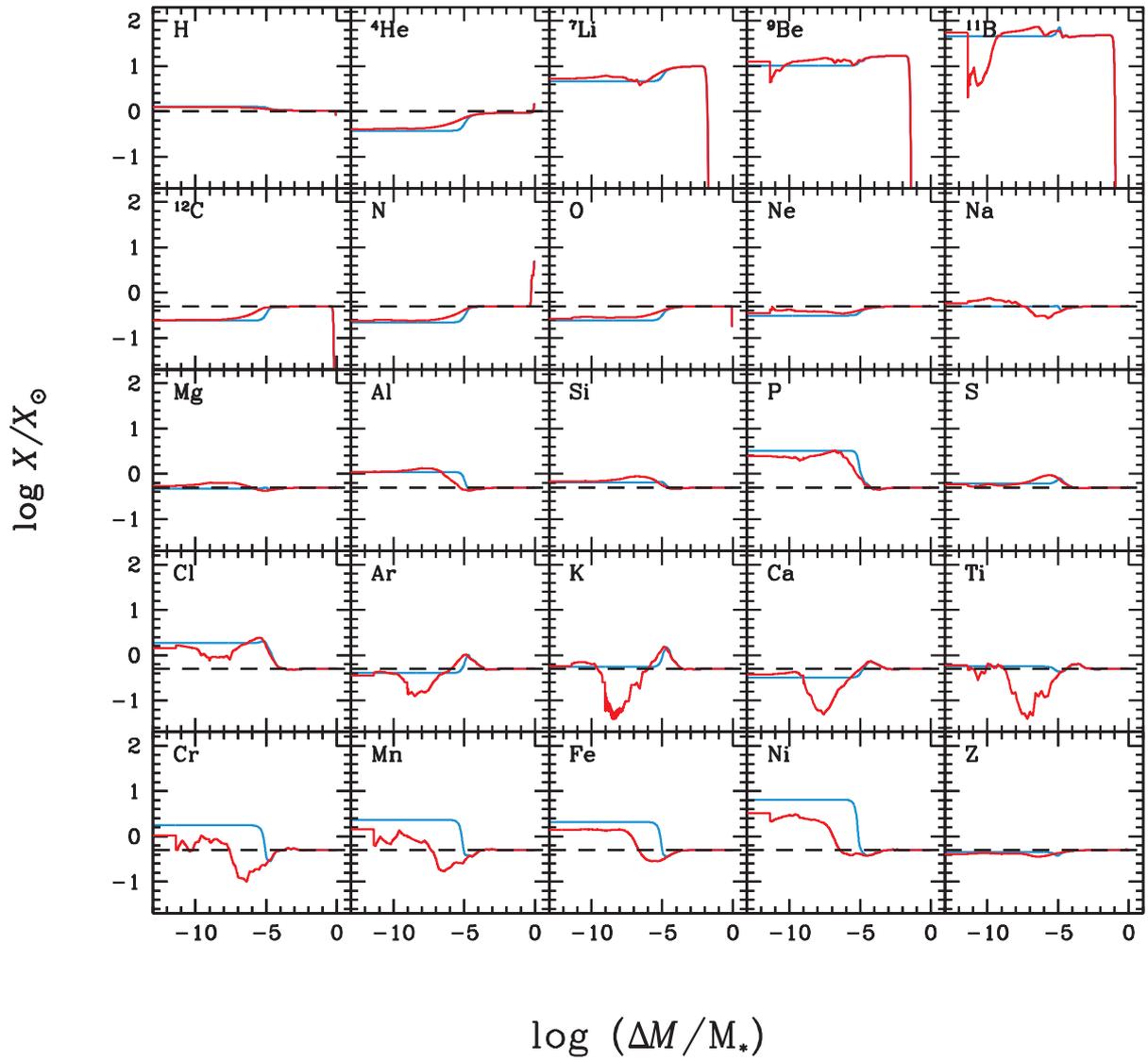}
      \caption{Mass fractions in the models with turbulence (blue curves) and mass loss (red curves). See the caption of Fig.\,\ref{fig:gR}.
      }
         \label{fig:internal_x}
   \end{figure*}   
\end{appendix}
\end{document}